\documentclass[a4paper]{jpconf}
\usepackage{graphicx}
\usepackage{cite}
\usepackage{mathrsfs}
\begin{document}
\title{Measurement of the electron drift velocity  for directional dark matter detectors}

\author{F.~Mayet$^1$, J.~Billard$^{2,3}$, 
G.~Bosson$^1$, O.~Bourrion$^1$, O.~Guillaudin$^1$, J.~Lamblin$^1$, J.~P.~Richer$^1$, Q.~Riffard$^1$, D.~Santos$^1$, 
F.~J.~Iguaz$^4$, L.~Lebreton$^5$, D.~Maire$^5$}

\address{$^1$Laboratoire de Physique Subatomique et de Cosmologie,
Universit\'e Joseph Fourier Grenoble 1, 
  CNRS/IN2P3, Institut Polytechnique de Grenoble, 
  53, rue des Martyrs, Grenoble, France}
\address{$^2$ Department of Physics, 
Massachusetts Institute of Technology, Cambridge, MA 02139, USA}
\address{$^3$ MIT Kavli Institute for Astrophysics and Space Research, Massachusetts Institute of Technology, Cambridge, MA 02139, USA}
\address{$^4$ Univ. de Zaragoza - Dep. de Fisica Teorica, Pedro Cerbuna 12, E-50009 Zaragoza, Spain}
\address{$^5$ LMDN, IRSN Cadarache,
13115 Saint-Paul-Lez-Durance, France}

\ead{mayet@lpsc.in2p3.fr}

\begin{abstract}
Three-dimensional track reconstruction is a key issue for directional Dark Matter detection. It requires a precise knowledge of the
electron drift velocity. Magboltz simulations are known to give a good evaluation of this parameter. However, large TPC operated underground
on long time scale may be characterized by an effective electron drift velocity that may differ from the value evaluated by simulation. {\it In situ}  measurement
of this key parameter is hence a way to avoid bias in the 3D track reconstruction. We present a dedicated method 
for the measurement of the electron drift velocity with the MIMAC detector. It is tested on two gas mixtures :  
$\rm CF_4$ and $\rm CF_4+CHF_3$. 
We also show that   adding $\rm CHF_3$ allows us to lower the electron drift velocity 
while keeping almost the same Fluorine content of the gas mixture.
\end{abstract}

\section{Introduction}
Directional detection of galactic Dark Matter offers
a unique opportunity to identify Weakly Interacting Massive Particle (WIMP) events as
such~\cite{spergel,morgan1,morgan2,green1,green2,green.disco,billard.disco,billard.exclusion,billard.ident,billard.profile,Billard:2012qu,Alves:2012ay,albornoz}.
This new   search strategy  requires  the simultaneous measurement of the recoil energy   and the direction of the 
3D track  of  low energy recoils. There is a worldwide effort toward the development 
of a large TPC (Time projection
Chamber) devoted to directional detection \cite{white,cygnus2011}. All current 
projects \cite{dmtpc,drift,d3,mimac,newage} face common challenges amongst which 3D track reconstruction \cite{billard.track} 
is the major one.\\
It is compulsory to have a precise knowledge of the electron transport  properties 
in the gas mixture used as a sensitive medium for  the TPC \cite{caldwell}, in particular the 
electron drift velocity. Indeed, primary electrons, created along the recoil trajectory, are used to retrieve the recoil track in the TPC. 
In particular, for the MIMAC project \cite{mimac}, 
the measurement of the third dimension,  along the electric field, is achieved thanks to a sampling of 
the primary electron cloud. Large TPC operated underground on long time scale may be characterized by an effective electron drift velocity that may differ from the value evaluated by Magboltz simulation \cite{magboltz}, due to {\it e.g.} impurities, field inhomogeneities, long drift distances. 
{\it In situ}  measurement of this key parameter is hence a way 
to avoid bias in the 3D track reconstruction.\\
Standard measurements of   the electron drift velocity are done thanks to photo-electrons generated by a  $\rm N_2$ laser
 \cite{schmidt1,schmidt2,Colas.drift}, by measuring  the  time difference between the UV emission time and the electron arrival time on the anode. In order to have a precise measurement,  
small drift spaces are  used \cite{Colas.drift}, $\mathcal{O}$(10) mm, ensuring that the electric field remains homogeneous. 
This method is not suited for large drift spaces and {\it in situ} measurement allows us to retrieve  
the effective drift velocity that electrons are encountering in the  Dark Matter detector.\\
We present a dedicated method for an {\it in situ}  measurement of electron drift velocity with the MIMAC detector 
\cite{mimac},  using  a collimated $\alpha$ source, together with a maximum likelihood 
method associated to a modelisation of the signal induced on the grid.

\section{Experimental setup}
\label{sec:derive:dispositif}
This electron drift velocity measurement is done directly with the MIMAC detector \cite{mimac}.
The primary electron-ion pairs produced by a nuclear recoil in the MIMAC chamber are detected by drifting 
the primary electrons to the grid of a bulk Micromegas \cite{iguaz} and 
producing the avalanche  in a very thin gap 256 $\mu$m.  The MIMAC prototype $\mu$TPC is composed of a pixelized anode featuring 2
orthogonal series of 256 strips of pixels (X and Y) and a micromesh grid separating  the amplification (grid to anode) and the drift space (cathode to grid). Each strip of
pixels is monitored by a current preamplifier and the fired pixel coordinate is obtained by using the
coincidence between the X and Y strips (the pixel pitch is 424 $\mu$m).
In order to reconstruct the third dimension of the recoil,   along the drift axis, 
 a self-triggered electronics has been developed \cite{Richer:2011pe,Bourrion:2011vk}. 
 It allows us to perform the anode sampling at a frequency of 50 MHz. Hence,  the effective electron drift velocity need to be 
 known, which is the goal of this paper.\\ 
For the  electron drift velocity measurement with the MIMAC detector,  we use a collimated $\alpha$ source 
($E_{\alpha} = 5.478$ MeV from $^{241}$Am), positioned on the cathode and facing the anode. 
As the energy loss  is about 3 MeV in the TPC ($50$~mbar of CF$_4$), the $\alpha$ particle is expected to 
cross the whole drift space (17.7 cm)
as well as the 256 $\mu$m amplification space. Since the $\alpha$ particle velocity is much greater than the electron drift velocity by about 2 orders of magnitude, the $\alpha$ arrival
time on the anode is taken as the starting time of primary electrons at the cathode.
The usual MIMAC amplification field has been lowered by a factor of 1.5, 
the $\alpha$ ionisation energy is around 3 MeV, much bigger than for low energy nuclear recoil.\\
In order to recover a non biased estimate of the drift velocity, we have measured the time dependent charge collection profile. The rise time of the charge sensitive preamplifier (when charges are injected on the grid), 
 is about 400 ns. This is much lower than the expected collection time of the primary electrons coming from the alpha tracks for the different gases and drift field considered hereafter, which are estimated using 
Magboltz to be between 1.5 $\mu$s and 17 $\mu$s for a pure CF$_4$ gas and a CF$_4$ + CHF$_3$ gas mixture 
respectively. Hence, we should be able to measure an accurate charge collection profile for each alpha track observed.

\begin{figure}[t]
\begin{center}
\includegraphics[scale=0.39,angle=0]{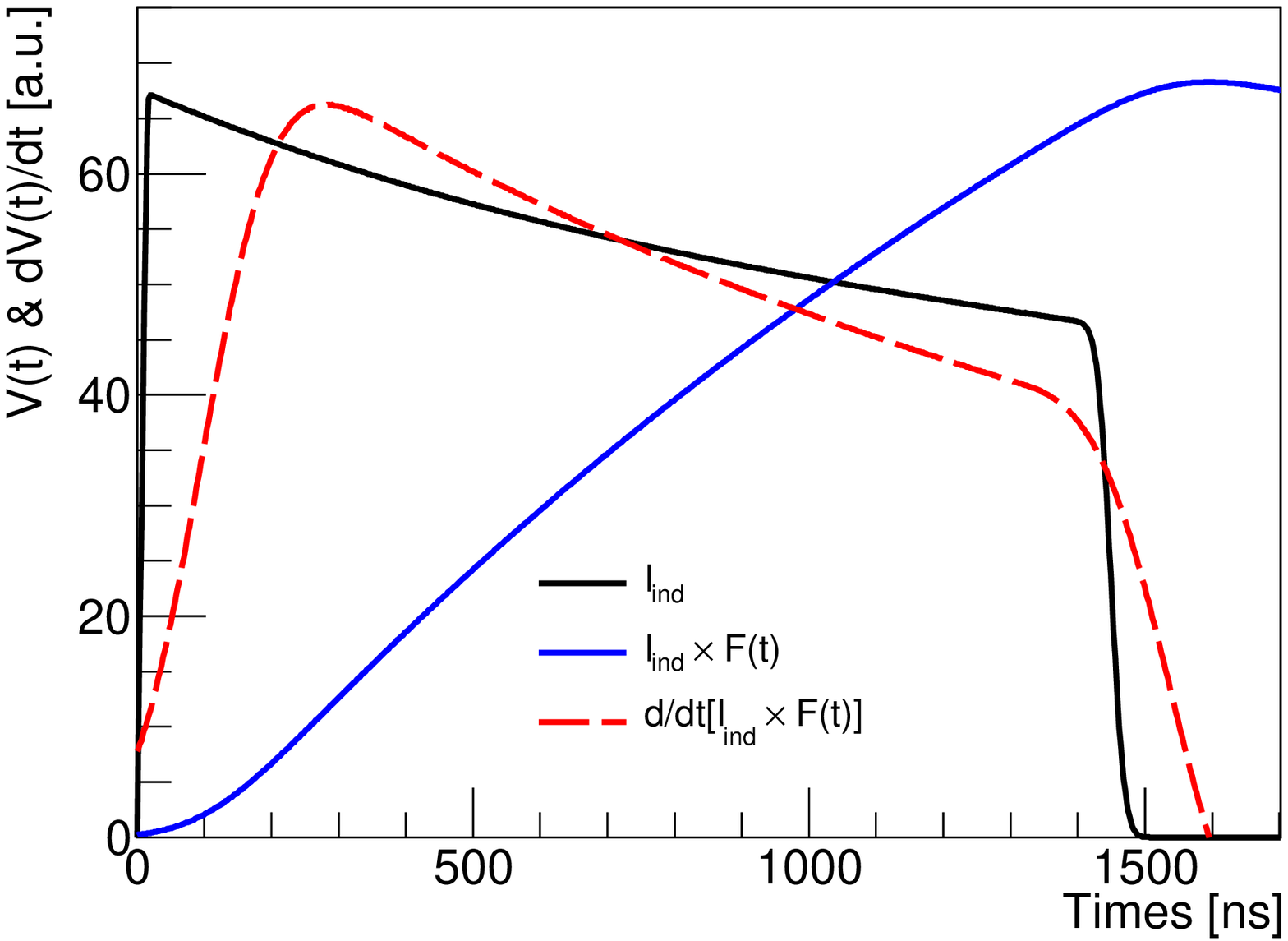}
\includegraphics[scale=0.39,angle=0]{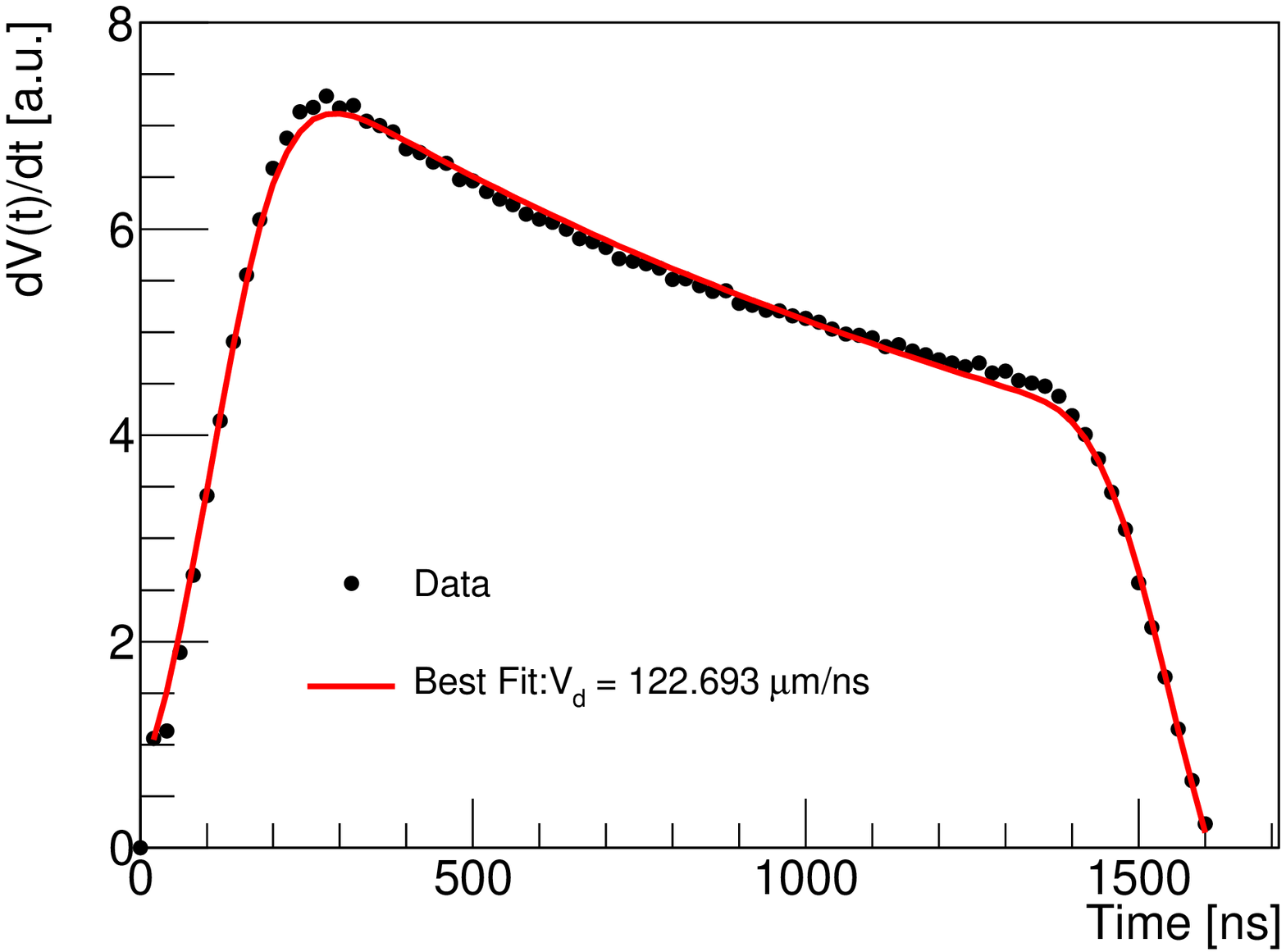}
\caption{\it (Left) Theoretical signal $V_{\rm th}(t)$ (blue curve) from the convolution of the current induced on the grid $I_{\rm ind}(t)$ (blaks curve) and the 
transfer function of the charge preamplifier $F(t)$. The dashed red curve is the time derivative  $V_{\rm th}'(t)$ of the signal. (Right) Mean profile (black point) and  best fit (red curve) for the  time derivate  of the signal 
$V^\prime(t)$. The measurement has been done for a pure  CF$_4$ gas  at $50$~mbar, with 
$E_d = 137.9$ V/cm and $E_a = 14.5$ kV/cm.} 
\label{fig:BestFit}
\end{center}
\end{figure}
\section{A likelihood-based data analysis strategy}
\label{sec:derive:vraisemblance} 
Straightforward data analysis strategies would not allow us to estimate the electron drift velocity without bias. For instance, one may evaluate the electron drift time, corresponding to a given drift length,  with the charge preamplifier, 
connected to the grid, by measuring the time between the maximum and the minimum. In this case, a significant lengthening of the output signal is expected, due to the readout time constant. This leads to an
under-estimation of $v_d$. The information contained in the 3D track may also be used through the use of the time difference between the first and the last spatial coincidence. While being almost not delayed by the electronic readout, this estimation is expected to depend  
heavily on the amplification electric field (the gain).  Indeed, the probability to have a 
spatial coincidence depends on the number of electrons contained in a given time sample and hence on the amplification gain.\\
To avoid bias due to electron diffusion, ion collection time and electronic readouts, we propose a data analysis 
strategy based on a profile likelihood method. It  requires a complete modeling of the 
signal, from the current induced on the grid to the measured signal  $V(t)$, in order to compare data and expected signals. 
The theoretical signal $V_{\rm th}(t)$ is  given by \cite{billard.drift}
\begin{equation}
V_{\rm th}(t) \propto \int\int\int \frac{dE}{dt}(t - \xi) \times Q_{\rm ion}(\xi - \tau) \times g_{\rm diff}(\tau -T) \times F(T)\ dT d\tau d\xi
\end{equation}
where  : 
\begin{itemize}
\item $dE/dt$ is the collection of primary electrons on the anode, {\it i.e.} the time projection, for a given electron drift velocity $v_d$, of the ionisation energy loss  $dE/dx$ simulated  with Geant 4 \cite{geant4}.
\item $Q_{\rm ion}(t)$ is the ion-induced current, approximated by a gate function of width $\Delta t_{\rm ion} =  \epsilon / v_{d_{\rm ion}}$, where $v_{d_{\rm ion}}$ is 
the drift velocity of ions in the amplification space and 
$\epsilon$ is the Micromegas amplification width  (256 $\mu$m) \cite{iguaz}. 
\item  $g_{\rm diff}(t)$ is a gaussian distribution accounting for the electron longitudinal diffusion, with a standard deviation 
  $\sigma_l(t) = D_l\sqrt{v_d\times t}$, where $D_l$ is the longitudinal diffusion coefficient.
\item $F(t)$ is the measured charge sensitive preamplifier transfer function \cite{billard.drift}
\end{itemize}
Figure~\ref{fig:BestFit} presents the theoretical signal $V_{\rm th}(t)$ (blue curve) obtained from the convolution of the induced current on the grid $I_{\rm ind}(t)$ (black curve) and the 
transfer function of the charge preamplifier $F(t)$  \cite{billard.drift}. The time derivate  
$V_{\rm th}'(t)$ of the signal $V_{\rm th}(t)$ is also presented (dashed curve). 
For this example,   the following parameter values are used : 
$D_l = 440 \ \mu {\rm m/\sqrt{cm}}$, $v_{d_{\rm ion}} \sim 8 \ \mu {\rm m/ns}$ and 
$v_d = 122 \ \mu {\rm m/ns}$.  In the following, 
the theoretical signal   $V_{\rm th}(t;v_d,v_{\rm ion},D_l)$ is the adjusting model used in the 
profile likelihood method that allows us to estimate the uncertainty on the electron  drift  velocity. 
This way  the uncertainties on all other parameters are accounted for, leading to a the profiled likelihood function that is larger 
($\rm \sim 40 \ \%$) than the standard one. Including systematics from the experimental setup, such as the length of the chamber
and the homogeneity of the electric field along the drift space, the uncertainty  
on the electron drift velocity is evaluated at the level of 1\%.\\
For each setup configuration, we measure   $\sim 500$ $\alpha$ tracks and we evaluate a mean profile $\bar{V}(t)$ that is being adjusted by the  
 signal model $V_{\rm th}(t;v_d,v_{\rm ion},D_l)$. However, evaluating the likelihood function as a product of likelihoods requires that each $V(t_i)$ are independent of 
each other. This is obviously not the case as the  $V(t)$ signal corresponds to an integration of the induced current and  we have checked that the correlation matrix 
is highly non diagonal. To avoid a diagonalization of the covariance matrix,  the time derivate of
the signal, $V^\prime(t)$, is used instead. The likelihood function can then be written as the product of the likelihoods associated with 
each value of  $\bar{V^\prime}(t_i)$. It reads as 
\begin{equation}
\mathscr{L}(v_d,v_{\rm ion},D_l,\delta t, A) = \exp\left(-\frac{1}{2}\sum_{i=1}^{N_t}\left[\frac{A\times V_{\rm th}'(t_i-\delta t;v_d,v_{\rm ion},D_l) - \bar{V}'(t_i)}
{\sigma_{\bar{V}'(t_i)}}   \right]^2\right)
\end{equation}
where $\delta t$ et $A$ are adjusting parameters, to enable a time and amplitude shift  
between the data and the
adjusting model.  $\bar{V^\prime}(t)$ is the mean profile value, $\sigma_{\bar{V^\prime}'(t)}$ its statistical standard deviation and $N_t$ is the number of time samples.
Note that the four nuisance parameters are associated with flat and non informative prior distributions. 

\section{Experimental results}
\label{sec:results}

\begin{figure}[t]
\begin{center}
\includegraphics[scale=0.39,angle=0]{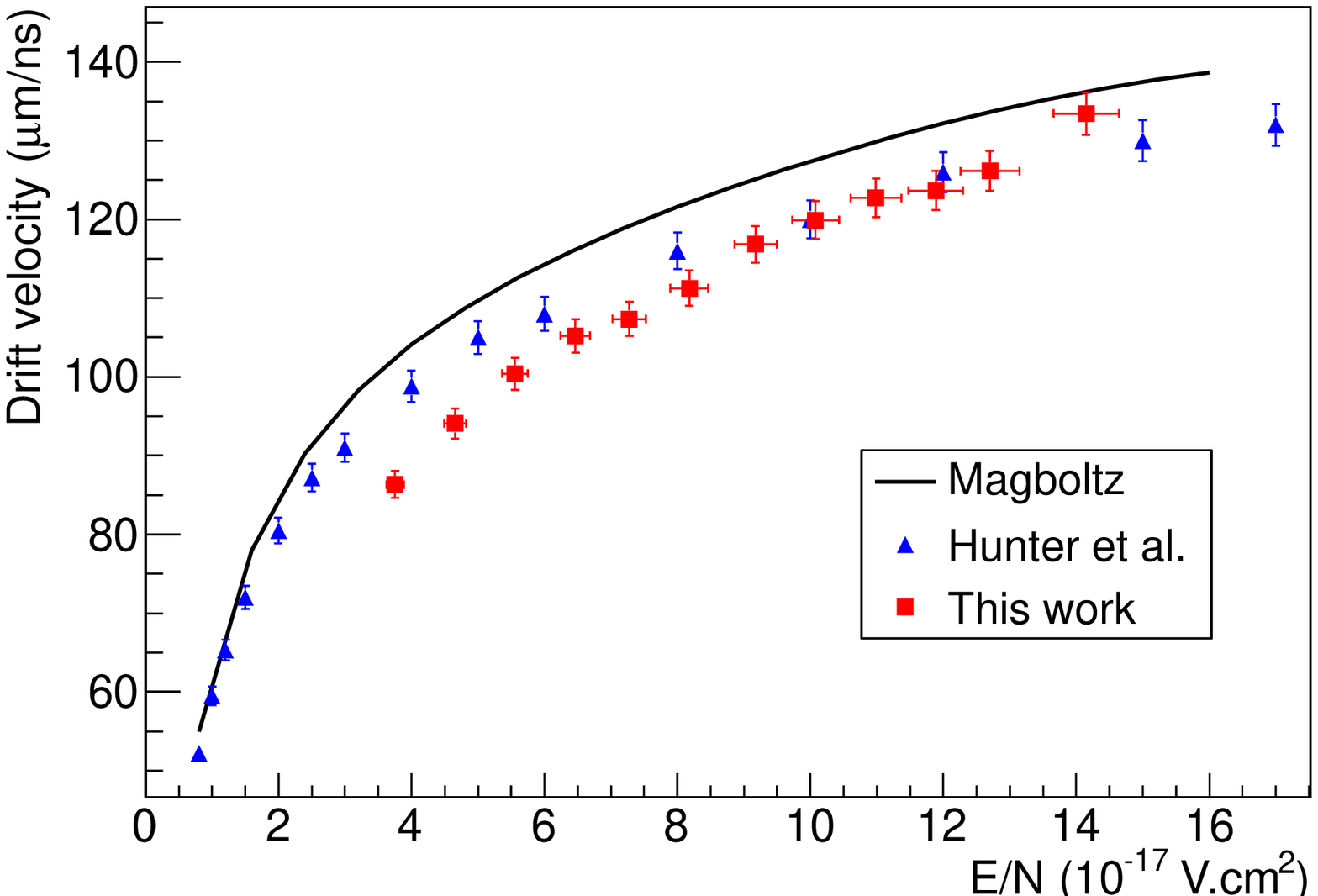}
\includegraphics[scale=0.39,angle=0]{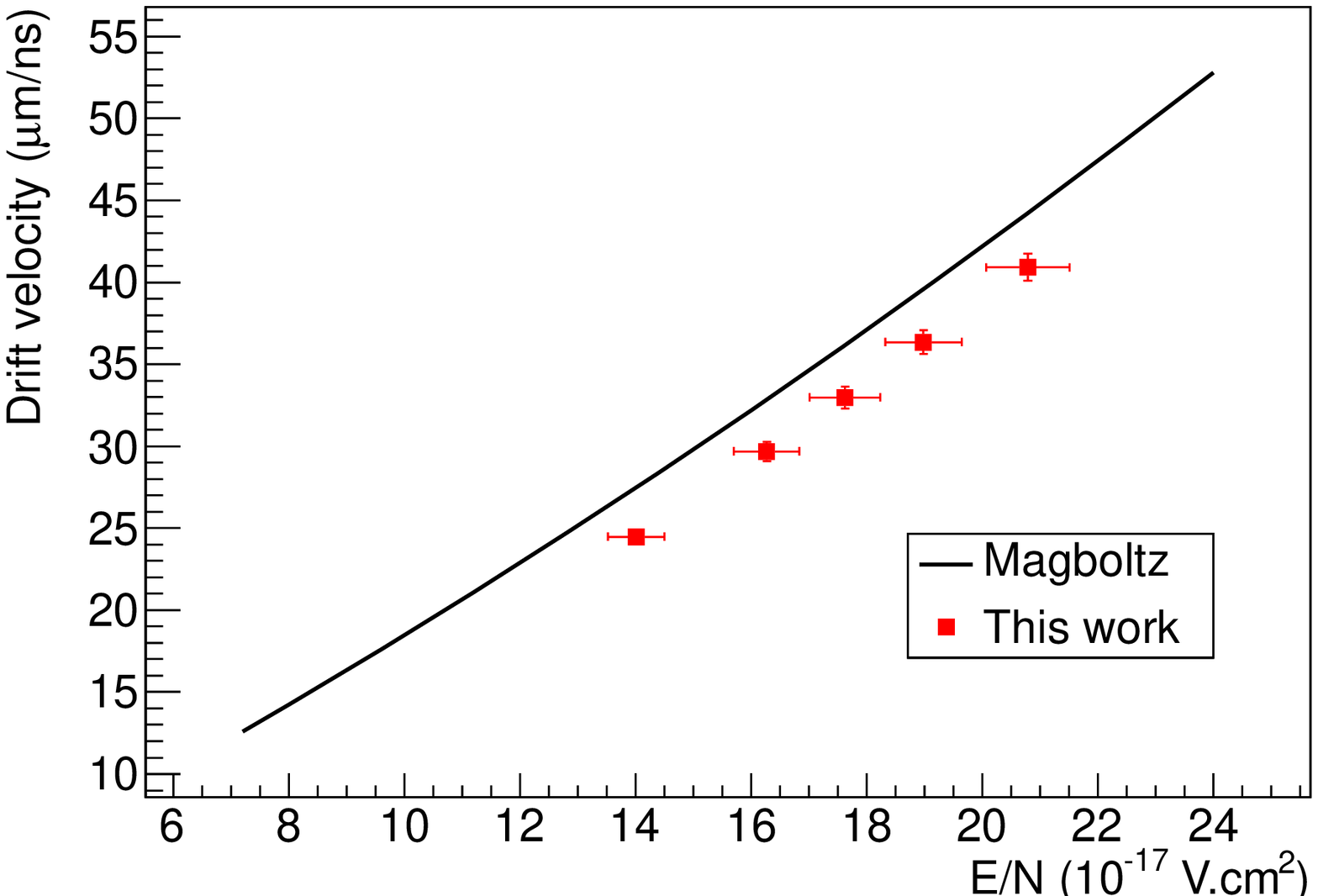} 
\caption{\it Electron drift velocity  $v_d$ ($\mu {\rm m/ns}$) 
 in  CF$_4$ (left) and  in a 70\% CF$_4$ + 30\% CHF$_3$ gas mixture (right) as a function of $E/N  \ (10^{-17}  \ {\rm Vcm^{2}})$,  at 
 $50$~mbar and for an
amplification field  $E_a = 15.6$ kV/cm for CF$_4$ + CHF$_3$  and $E_a = 14.5$ kV/cm for CF$_4$. The Magboltz simulation
(black line) and previous CF$_4$  data (blue triangle) \cite{hunter} are also presented.}
\label{fig:DriftVelocityCF4CHF3}
\end{center}
\end{figure}

This method for {\it in situ} electron drift velocity measurement   has been applied  to two gas mixtures 
that might be used for directional Dark Matter detection: pure $\rm CF_4$ and  $\rm  CF_4 + CHF_3$. 
To examplify, we present on figure~\ref{fig:BestFit} (right) the mean profile (black point) and  best fit (red curve) for the  time derivate  of the signal 
$V^\prime(t)$, for  pure $\rm CF_4$ at $50$~mbar and a drift field of $\rm E_d = $ 137.29 V/cm.  The adjustment is excellent, in
particular in the regions of interest for the estimation of the electron  drift  
velocity, {\it i.e.} rising and falling part of the  mean profile $\bar{V^\prime}(t)$. Small differences between the  fit 
and the data    in the central region ($300 \ {\rm ns} \leq t \leq 1200 \ {\rm ns}$) are due to a lack of accuracy in the estimation of the decreasing part of the preamplifier transfer function.\\
Figure~\ref{fig:DriftVelocityCF4CHF3} (left) presents the electron drift velocity measurement in a pure CF$_4$ gas at $50$~mbar, for an
amplification field  $E_a = 14.5$ kV/cm and a drift field $E_d$ ranging between 50 V/cm and 175 V/cm. 
The results obtained from a  Magboltz simulation \cite{magboltz} are also presented and compared with 
previous experimental data \cite{hunter}. Fig.~\ref{fig:DriftVelocityCF4CHF3} (right) presents the electron drift velocity measurement in a 70\% CF$_4$ + 30\% CHF$_3$ 
gas mixture at $50$~mbar, for an amplification field  $E_a = 15.6$ kV/cm. To our knowledge there is no other experimental data with this
gas mixture.  As expected,
 there are discrepancies with the Magboltz simulation as the drift velocity measured accounts for real, but unknown, 
experimental conditions (impurities, field inhomogeneities, ...) and corresponds to a long drift distance. 
Note that the electron drift velocity in a 70\% CF$_4$ + 30 \% CHF$_3$ gas mixture   is about 5 times lower than in the pure 
$\rm CF_4$ case. This is a key point for directional Dark Matter as the track sampling along the drift field 
will be improved  while keeping almost the same 
Fluorine content of the gas mixture. Fluorine is indeed a golden target \cite{white,cygnus2011} for spin-dependent Dark Matter search as the spin content is
dominated by the unpaired proton \cite{albornoz}. In the case of the MIMAC detector, a fraction of 30\% was found to be
adequate as it allows to significantly enhance the 3D track reconstruction while
conserving sufficiently dense primary electron clouds needed for a high nuclear recoil track detection efficiency.

\section{Conclusion}
We have presented a  new method for {\it in situ} electron drift velocity measurement, 
using an alpha source and a profile likelihood analysis based on the modeling of the signal induced on the grid. 
In particular, we have shown that such analysis allows us to avoid
bias due to {\it e.g.}  electron diffusion, ion collection time and electronic readout.
Hence, the effective electron drift velocity,  {\it i.e.} in the whole drift space, is measured and it corresponds to the real 
experimental conditions of the upcoming MIMAC directional detector. 
We also suggest to add $\rm CHF_3$ to the standard $\rm CF_4$ gas
used for directional detection as it allows us to lower the electron drift velocity 
while keeping almost the same Fluorine content of the gas mixture. 
 


\section*{References}


\end{document}